# Machine Learning for Optical Scanning Probe Nanoscopy

Xinzhong Chen[1][†], Suheng Xu[2][†], Sara Shabani[2], Yueqi Zhao[3], Matthew Fu[2], Andrew J. Millis[2], Michael M. Fogler[3], Abhay N. Pasupathy[2], Mengkun Liu[1,4*], D. N. Basov[2*]

[1]Department of Physics and Astronomy, Stony Brook University, Stony Brook, New York 11794, USA

[2]Department of Physics, Columbia University, New York, New York 10027, USA

3Department of Physics, University of California at San Diego, La Jolla, California 92093-0319, USA

[4]National Synchrotron Light Source II, Brookhaven National Laboratory, Upton, New York 11973, USA

†co-first authors

* mengkun.liu@stonybrook.edu   db3056@columbia.edu

**The ability to perform nanometer-scale optical imaging and spectroscopy is key to deciphering the low-energy effects in quantum materials, as well as vibrational fingerprints in planetary and extraterrestrial particles, catalytic substances, and aqueous biological samples. The scattering-type scanning near-field optical microscopy (s-SNOM) technique has recently spread to many research fields and enabled notable discoveries. In this brief perspective, we show that the s-SNOM, together with scanning probe research in general, can benefit in many ways from artificial intelligence (AI) and machine learning (ML) algorithms. We show that, with the help of AI- and ML-enhanced data acquisition and analysis, scanning probe optical nanoscopy is poised to become more efficient, accurate, and intelligent.**

## Introduction

Machine learning (ML) and artificial intelligence (AI) are revolutionizing a wide range of industries and sciences at an unprecedented pace. A properly trained artificial intelligence agent is capable of performing a series of complex tasks ranging from comprehending information from human speech to maneuvering autonomous vehicles in busy traffic. Although numerous ML/AI concepts and algorithms such as the artificial neural network (NN) that are vastly popular today were developed decades ago, efficient training of deep ML models has only been enabled by the recent advances in computer architecture and hardware. At the same time, a tremendous amount of data, ranging from customers' shopping preferences on an e-commerce site to particle trajectories from the Large Hadron Collider, are generated and collected on a daily basis. The sheer volume of the raw data urgently calls for a systematic way of extracting the information hidden within, which is exactly where ML and AI shine. Historically stemming from computer and data science, ML has quickly found legions of applications in physical sciences research thanks to its unique ability to mine nontrivial information from data[1]. Capitalizing on the availability and accessibility of large volumes of data from research facilities and novel simulation platforms, ML provides a systematic approach to unveiling "hidden" information across many branches of physics [2–5].

The idea of using ML to study physics is sprouting across all its sub-disciplines [1], from exotic particle detection[2,3] to the identification of phases in quantum many-body systems[4,5]. This unprecedented bandwidth of ML utility stems from the fact that in many of the problems, the complexity is beyond the reach of generalized analytical solutions or simulations based on common computing approaches. ML is also no stranger to the field of optics/photonics[6–10]. For example, it has been shown that the design of functional metamaterials can be significantly accelerated by ML; traditionally, it is done by performing iterations of simulations and experiments in a trial-and-error fashion[11–18]. ML in this case improves not only the time efficiency but also the design accuracy. Furthermore, ML has also been applied to the emerging field of quantum optics for various classification, optimization, and regression tasks[19–21]. Advanced statistical tools and AI algorithms have also been implemented for image processing and pattern analysis in the field of scanning transmission electron microscopy (STEM) with various multimodal and hyperspectral microscopies[22–26]. On the other hand, scanning probe microscopy (SPM) techniques such as scanning tunneling microscopy (STM) and atomic force microscopy (AFM) have also been shown to benefit from ML in terms of scan acceleration, noise reduction, measurement automation, and physical property extraction[27–30]. However, at the emergent interface of optics and SPM – optical scanning probe microscopy such as scattering-type scanning near-field optical microscopy (s-SNOM)[31] – research regarding ML and AI is still in its infancy[32–34]. In this article, we discuss a handful of already implemented realizations of ML and AI on s-SNOM; we will augment this discussion by outlining specific potential applications where ML/AI are poised to empower progress. Since this is still a nascent field, this concise perspective is intended to be inspirational rather than comprehensive.

In Fig. 1, we identify three complementary utilities of ML particularly beneficial for nano-optical characterization -- supervised learning, unsupervised learning, and reinforcement learning. All three utilities can be employed independently, sequentially, or simultaneously to tackle different aspects of the s-SNOM experiments from data collection to data analysis. For example, a reinforcement learning agent can potentially be trained to perform automatic sample identification, optical optimization and alignment, and rapid scanning[35–37]. The acquired data can be processed by either supervised or unsupervised learning to extract the relevant material properties or new physical laws from the experimental observables[33,38]. In the following, we introduce several ideas and illustrate these ideas with concrete examples. Even though we present and evaluate the ML methods in the context of s-SNOM, these methods apply more broadly to a range of scanning probes or spectroscopy techniques such as tip-enhanced photoluminescence (TEPL)[39–43], tip-enhanced Raman scattering microscopy (TERS)[44–48] and scanning tunneling microscopy (STM)[30,49–54].

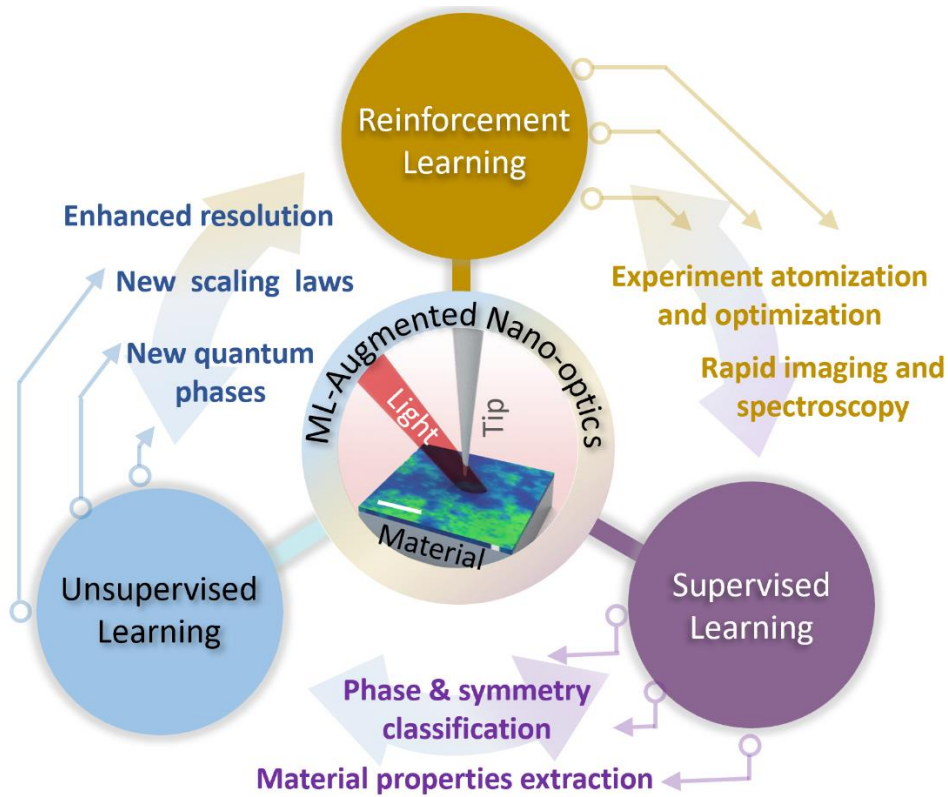

**Figure 1. Schematics illustration of different ML concepts applied to s-SNOM experiments from data collection to data analysis.** The image in the center visualizes a percolation process in the course of the metal-insulator transition in $VO_2$ thin film taken at ~10 μm wavelength of incident radiation[55]. Scale bar represents 500 nm.

**Machine learning for nano-spectroscopy**

We begin with a brief review of the current mainstream data analysis efforts in optics and s-SNOM without the help of ML. A common initial goal of an optical measurement is the accurate extraction of the optical constants from observables obtained from experiments using such techniques as Fourier-transform infrared spectroscopy (FTIR)[56,57] terahertz time-domain spectroscopy (THz TDS)[58–66], and many other complementary methods. Due to the well-understood physics behind the light-matter interaction, the extractions are often straightforward. However, the diffraction limit imposes a fundamental constraint on the achievable spatial resolution and also restricts access to the optical response function away from zero momentum[67]. To circumvent these shortcomings, tip-based optical scanning probe techniques such as s-SNOM, where the spatial resolution is practically limited only by the size of the tip apex instead of the wavelength, quickly emerge as an attractive alternative and advancement. At the same time, the analysis of s-SNOM data needs to take into account the complex electrodynamic interplay between the tip, the incident light, and the sample in the near-field regime, which imposes formidable obstacles in establishing simple procedures for the extraction of relevant physical quantities. So far, most of the extraction schemes have been heavily based on physics-driven analytical modeling. That approach relies on a model that accurately describes the electrodynamics of the tip-sample interaction. Examples of such models include the point-dipole model[68,69], the finite-dipole model (FDM)[70], the lightning-rod model[71], and their variants[72–76]. In these models, simplified geometries such as a sphere or an elongated spheroid are typically employed to approximate the tip.

In s-SNOM measurements, an important procedure is to drive the tip to harmonic oscillation at the cantilever's mechanical resonance frequency (tapping mode AFM). This oscillation modulates the near-field interaction. Subsequently, demodulation of the detector signal at higher harmonics of the tip oscillation frequency by a lock-in amplifier isolates the near-field signal from the overwhelmingly large background signal due to sample surface and cantilever scattering[77,78]. The background signal is assumed to be insensitive to the tip taping because the oscillation amplitude, in the orders of tens of nanometers, is much smaller than the free-space wavelength. Optical constant extraction from s-SNOM experiments based on analytical models has been successfully demonstrated[71,79–84]. For example, as shown in Fig. 2(a), both the optical constant and the layer thickness can be simultaneously recovered from the experimental near-field signals demodulated at various harmonics of the tip tapping frequency[79,82]. This is because the near-field signals from different demodulation orders encode material properties at different probing depths [82,85] – the higher the order, the shallower the probing depth. Similarly, the dielectric function of the sample can be determined from near-field spectroscopy (nano-FTIR) in a broad spectral range measurement by using a numerical inversion protocol based on physical modeling (Fig. 2(b)[80]).

In addition to monochromatic nano-imaging and broadband nano-spectroscopy, s-SNOM provides additional important measurement modalities such as time-resolved pump-probe spectroscopy[86–94] and hyperspectral imaging[95–98]. In the near-field pump-probe experiments, time-resolved nano-spectra or images can be obtained to yield carrier dynamics with femtosecond resolution, as shown in the example in Fig. 2(c). Hyperspectral imaging, on the other hand, requires a series of nano-spectra taken in close proximity to each other within a micron scale region. As shown in Fig. 2(d), a hyperspectral image is taken normal to the sample edge of hBN on $SiO_2$ at the mid-IR frequency range[99], which can yield the energy-momentum dispersion relation of hBN by means of profile fitting[98,100–104]. Furthermore, the clear technical trend points to multimodal imaging via a combination of s-SNOM and other optical or non-optical SPM techniques[105–108]. For example, electronic and magnetic correlation imaging using s-SNOM and magnetic force microscopy (MFM) have led to intriguing discoveries in manganites[107,109]. Other optical SPM techniques such as the TEPL and the TERS have demonstrated molecular level spatial resolution[110,111] and can be co-operated with s-SNOM [Fig. 2(e)]. For data analysis, however, all the above examples benefit from analytical solutions or simulations of the tip-sample interactions using simple toy models.

Quantitative response function extraction in nano-optics was demonstrated for a variety of materials[71,79–84]. Many interesting material systems such as polar crystals and doped semiconductors exhibit strong resonances linked to various excitations including phonons and plasmons in the technologically relevant IR and THz regimes[112]. An oversimplified model prediction is known to systematically bias the outcome compared to experiments, especially in the strong resonance regime[112]. Notably, the details of the tip geometry and other experimental conditions remain largely overlooked in the current simplified physical models. Furthermore, other factors such as the topography and size of the sample are known to produce near-field signal contrast that is not intrinsically optical[113,114]. For these reasons, rigorous mathematical description of the tip-sample interaction that includes all the details is unrealistic and significant errors are likely to persist.

Recent advances in ML-based practices point to an interesting alternative to conventional physical modeling. Within the data-driven approaches, the quantitative connections between various physical quantities are established based on the data alone, eliminating the external bias caused by imperfect modeling or other approximations. Provided a sufficient amount of high-quality training data with proper labels is available, supervised learning is the natural technique of choice. For example, training data can be collected using nano-FTIR on a wide range of common materials with well-characterized optical properties. An NN can be trained using these nano-spectra and the known dielectric functions of the materials. The trained NN establishes a functional mapping that connects the near-field spectroscopic response and the optical constants, similar to an analytical model[32].

Importantly, the analytical models can serve as an additional information source for the ML. It has been demonstrated that an NN infused with an analytical model outperforms regular NNs in terms of fidelity for the determination of the dielectric functions from nano-spectroscopy data[33] (Fig.3a). The approach is also shown to be superior to the analytical model alone and requires a smaller amount of input data compared to the purely data-driven NN. Intuitively, this is because the added model serves as a guide for the NN, thereby reducing the complexity of the function the NN attempts to learn. Such an approach is especially valuable when high-quality experimental data are either scarce or sparse.

As shown schematically in Fig. 3(b), to fully decipher an s-SNOM spectrum on samples with complex heterogeneity such as thin layers, irregular geometry, or anisotropy, a large number of experimental parameters is needed. These parameters include illumination frequency and angle, tip geometry, tapping amplitude, and many others. To extract the intrinsic material properties of interest, such as the full dielectric tensor and the layer thickness, all the relevant experimental parameters need to be taken into account, in principle. This multi-dimensional parameter space implies a massive volume of required training data. Fortunately, again, this can be combated by adding physical information into the ML model. Currently, there are several ways to augment the ML model with physical information -- for example, by enforcing certain governing differential equations that describe the system[14,115–119]. Physics inputs can also come from analytical models, numerical simulations, and first-principle calculations. It is important to note that although NN models are powerful in terms of versatility and expressiveness, there are potentially suitable ML models other than NN. Several ML regression models such as random forest or support vector machine can also be applied [120], the efficiency of which requires further exploration.

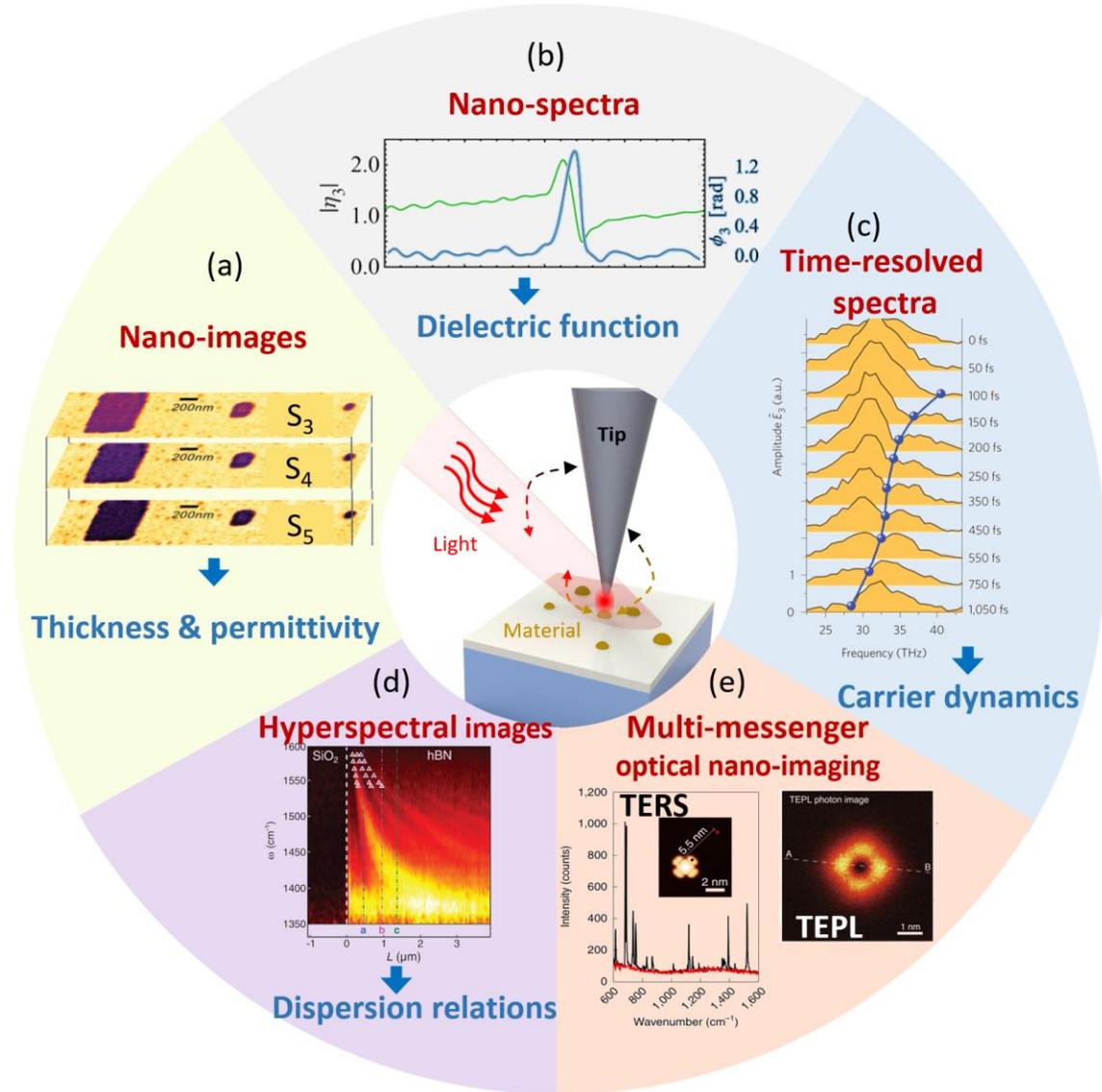

**Figure 2. Experimental observables and extracted parameters of s-SNOM nano-imaging/nano-spectroscopy along with other common optical SPM techniques.** These published works demonstrate state-of-the-art model-based data analysis efforts. (a) Complex near-field nano-images on PMMA nanostructures at different demodulation orders used to extract layer thickness and the dielectric function[79]. (b) Complex near-field nano-spectrum on PMMA, used to extract the frequency-dependent dielectric function[80]. (c) Time-resolved near-field spectra on an InAs nanowire pumped by near-IR pulse. The spectra can be used to extract time-dependent carrier dynamics[89]. (d) Near-field hyperspectral imaging of the hBN-SiO$_2$ boundary[99]. The Surface phonon polariton (SPhP) interference patterns are fitted to infer the frequency- and momentum-dispersion relations of SPhP. (e) Examples of TEPL and TRES data with molecular level resolution[110,111].

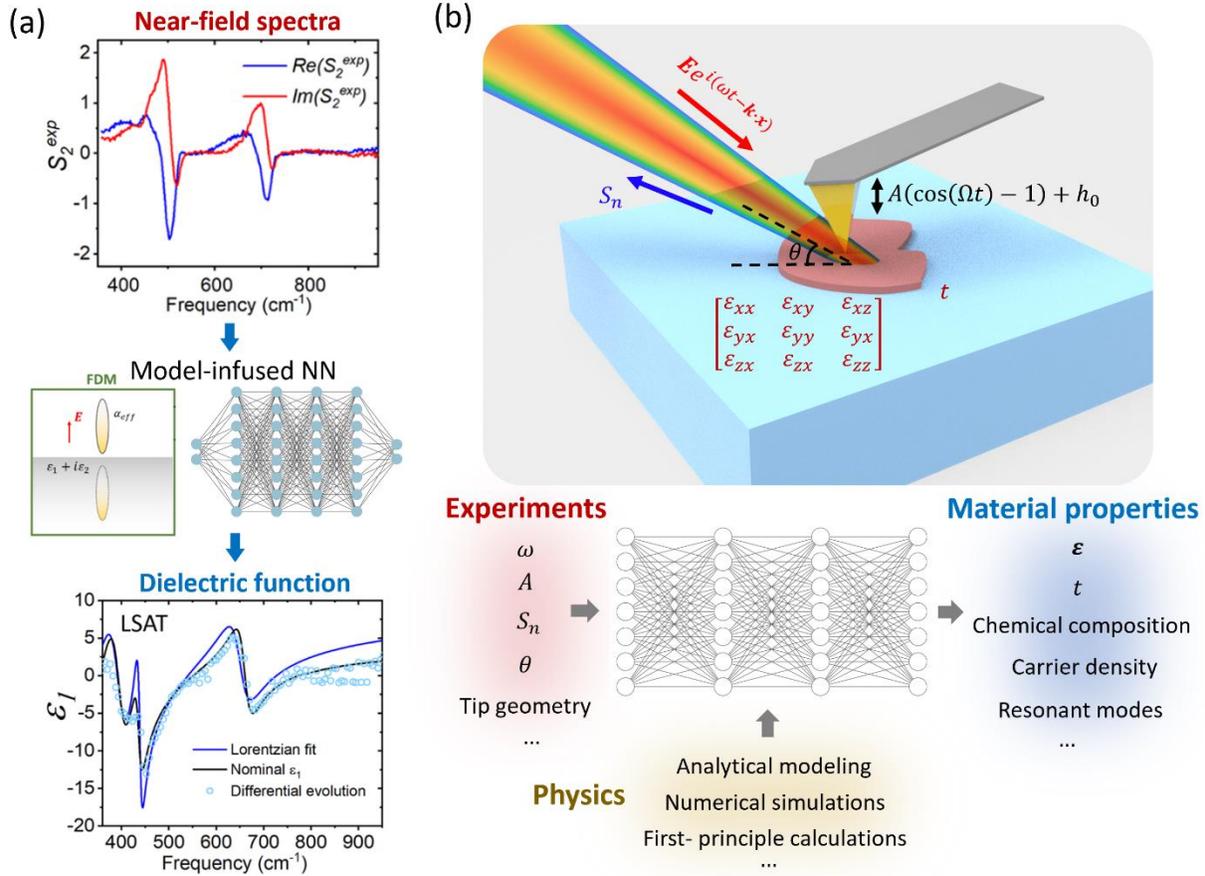

**Figure 3. Supervised learning for material property extraction.** (a) Experimental demonstration of the dielectric function of $(LaAlO_3)_{0.3}(Sr_2AlTaO6)_{0.7}$ (LSAT) extracted by the use of a physics-infused neural network that combines the FDM and the neural network[33]. (b) Top: schematics of the relevant experimental parameters and physical quantities. Bottom: We envision the ultimate ML model that quantitatively maps the experimental parameters/observables to the material-related quantities with the help of physical information from various sources.

### Machine learning for polariton imaging analysis

As mentioned above, collective electromagnetic modes can be effectively excited by the high in-plane momentum photons and get scattered by the AFM tip in the near-field. This leads directly to a particularly fruitful application of s-SNOM: the visualization of surface and bulk polaritons in van der Waals materials and the extraction of their linear or nonlinear response functions[72,98,99,104,121–135]. A representative polaritonic image is shown in Fig. 4(a), where the surface plasmon polaritons (SPPs) are visualized for the graphene/$RuCl_3$ interface[103]. The Fermi level and scattering rate of graphene can be extracted by fitting the wavelength and the quality factor of surface plasmon polaritons[136]. However, data postprocessing for polariton wave images is time-consuming and even more challenging on samples with high damping or nontrivial geometry. To accelerate data analysis, one can utilize the image recognition techniques developed in the field of ML using deep convolutional neural network (CNN)[137–140]. Because of its excellent performance with complex image feature extraction such as facial recognition, CNN prevails in physics and material science applications, especially in the field of microscopy and holography techniques[141–145]. The superiority of CNN for parameter extraction in SNOM images was recently demonstrated (Fig.4(a))[34]: trained with several thousand synthetic polariton wave images, the CNN simultaneously extracts the wavelengths and the quality factors from the polariton wave images in a millisecond timescale. The intermediate visualization from the feature extraction process reveals that the CNN maps out the interference fringes while ignoring the irrelevant features, e.g., the signal and noise on the substrate area, as is shown in the bottom part of Fig.4(a). The CNN-based parameter extraction not only shows great efficiency and fidelity but also demonstrates the potential capability of coping with more complex images and subtle features. It is also demonstrated that the CNN can extract multiple wavelengths for hyperbolic waveguide modes with better accuracy than the traditional Fourier transform[34]. Moreover, this approach can be applied to the polaritonic interference patterns near curved edges or in the polaritonic cavities, once the corresponding training data are included.

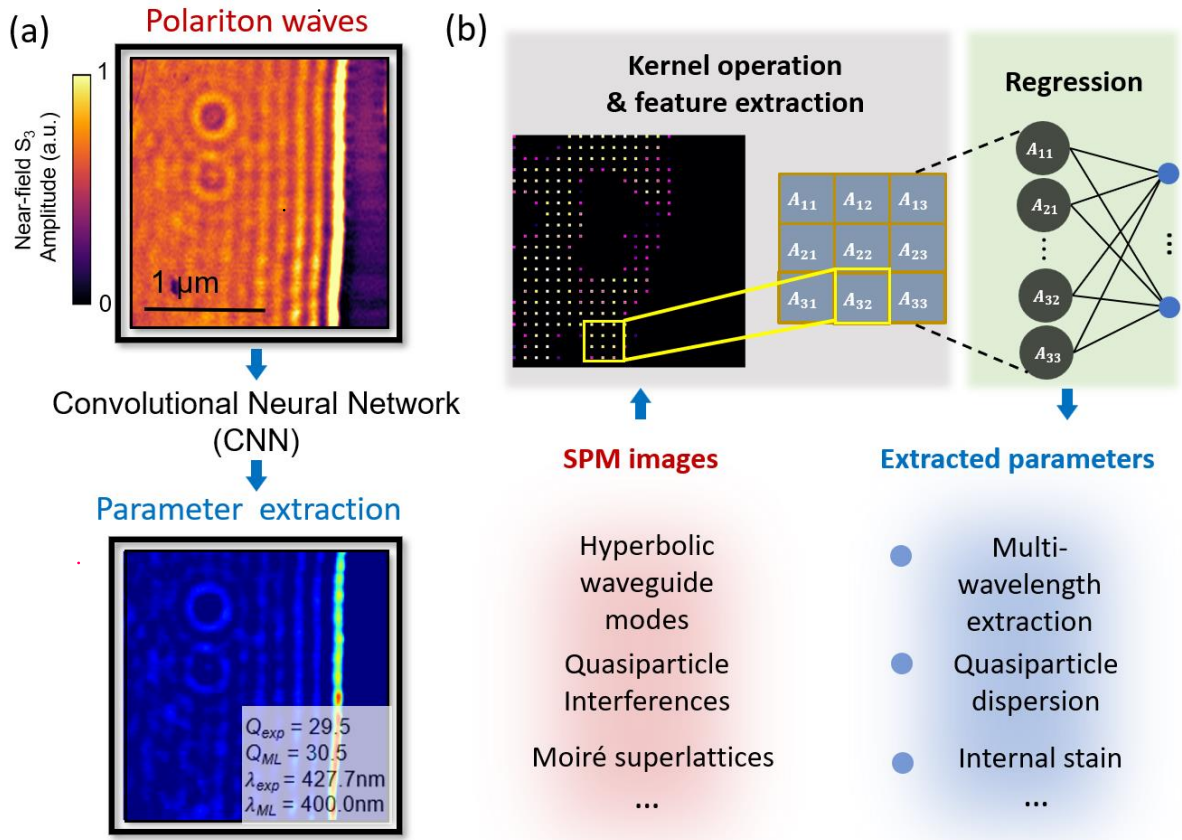

**Figure 4. Supervised nano-image analysis.** (a) Experimental polaritonic wave images and extraction results based on the convolutional neural network (CNN)[103]. The polariton wavelength and the quality factor are directly extracted from images in 150 ms[34]. (b) Top: schematics of CNN for SPM image parameter extraction. Bottom: Potential SPM image types for CNN parameter extraction and the corresponding extracted properties.

**Machine learning for multi-messenger nanoprobes including STM**

The ML algorithms developed for s-SNOM images are generic and are well suited for other SPM techniques, where nano- and micro-scale textures can be characterized by a few physical parameters. One example of the structure of CNN for SPM image parameter extraction is shown schematically in Fig.4(b). In the first part of the CNN, image features are extracted and distilled via multiple convolution layers and pooling layers. The images are shrunk to a few numbers for further regression in the fully connected layer. Trained with the properly labeled simulated or experimental dataset, a CNN can eventually extract the critical parameters from new SPM images. This approach has several foreseeable applications in SPM data analysis. Here we list a few plausible scenarios: 1) The dispersion of quasiparticles can be extracted by CNN from the quasiparticle interference images probed by scanning tunneling spectroscopy (STM) [146–148]. 2) In twistronics, large periodic superlattices are formed by twisting two layered materials [149,150]. CNN-based parameter extraction can be useful for evaluating the twisting angles, lattice parameters/internal strain, fermi-levels, or degree of electron localization associated with Moiré superlattice imaging via STM, AFM, s-SNOM, or MIM[106,108,151,152]. 3) A multi-messenger nanoprobe can be developed to yield simultaneous topography, local optical properties, and magnetic properties under external excitation[107]. These multi-modal imaging techniques yield high-dimensional data that are difficult to analyze with traditional methods. 2D or even higher-dimensional CNNs would play an important role in analyzing and discovering hidden patterns in those massive datasets of co-located images[153,154]. In the following, we present some specific examples of how ML is used to analyze STM images in the search of quasiparticle interference and atomic-scale periodic orders.

Scanning tunneling microscopy (STM) is a powerful local probe to visualize the electronic properties of surfaces with atomic resolution. Among the most common uses of this tool is as a probe of features of the nanoscale order such as charge density waves, which appear directly in imaging experiments[155]. A more advanced technique of data acquisition is to measure the local density of states as a function of energy, which gives rise to three-dimensional data sets in two spatial and one energy dimension. Additional dimensions such as temperature[156] and strain[157] have also been recently added to STM capabilities. Extracting relevant information from these multi-dimensional data sets can be a formidable task, due to the presence of aperiodicity arising from material defects, and the nontrivial interplay between local electronic structure and disorder.

A common analysis method in STM has been to perform Fourier transforms (FT) of real space images to extract periodicities present in the data. Such analyses have been useful for the detection of spatially ordered quantum phases[155] and the analysis of quasiparticle interference from defects in a material[147]. However, FT on aperiodic images such as scattering patterns around defects disregards fundamental information due to phase noise. Recently, ML algorithms based on techniques such as non-convex optimization have helped detect fundamental motifs in images[52]. Training artificial neural network (ANN) is an alternative method that has been used for the recognition of different hidden orders in hole-doped cuprate superconductors[143]. The developed ANN technique enables processing and identifying unidirectional lattice-commensurate and transitional symmetry-breaking (density wave) states. This supervised ML method holds great promise for identifying symmetry-breaking orders such as nematicity in strongly correlated materials, where even in the absence of sharp Bragg peaks, the ANN still successfully detects nematic order in $Bi_2Sr_2CaCu_2O_{8+x}$[53].

With the availability of newer ML techniques, the classes of STM problems that can be addressed using ML continue to grow rapidly. Deep learning networks (together with DFT) can be applied to detect and classify different types of defects and to provide insight into defect and materials functionalities[49,51]. In addition, deep learning infrastructure characterizes the molecular configuration in AFM[158] or STM images with structure prediction capability.

**Searching for new physics via unsupervised learning**

In previous sections, we discussed how ML and AI algorithms can help to process experimental data and extract physically relevant parameters mostly in a supervised learning fashion, where the output of one ML model is specified by humans and labeled training data are fed into the model. However, in many instances, the physical labels are unknown. For example, in complex quantum materials, revealing the interplay and competition between charge, spin, and lattice degrees of freedom often leads to nanoscale phase inhomogeneities[55,107,109,159–167]. It is hard to define one or several parameters that can fully characterize phase separation in the experimental image (see Fig.6(a)). Several studies have employed cluster statistics to quantify fractal geometrical dimensions and hence infer the criticality and universality of the phase transition in a "supervised manner." For example, after mapping the SPM images to Ising lattices, the universal classes of phase transition were determined by extracting geometrical critical exponents for the domains of the insulating and metallic phases[168–170].

The unsupervised learning approach is potentially promising for identifying new unknown trends, scaling laws, or hidden parameters. A recent study demonstrated that physical concepts such as conservation laws or degrees of freedom in quantum systems can be discovered directly from raw data[171]. A promising ML model -- autoencoder (AE) – may offer insights into the SPM data on strongly correlated quantum materials[171]. AE is a generative model that creates a low-dimensional latent parameter space and yields new images with one assigned point in latent space[172]. A simplified structure of AE is shown in Fig.6(b). A complete AE is composed of an encoder and a decoder operating in latent parameter space. In the encoder, the dimension of the input images is compressed down to several latent parameters. All the information from the original data is stored in the latent parameters and thus the decoder can regenerate the original data based on a few numbers. The difference between original data and generated data is minimized in the network training.

AE was first applied to the study of phase transition in 2017[173] and its use for studying phases of matter has exploded since then[174–179]. AE was also applied to the study of Ising lattices and showed that the latent parameter is exactly the order parameter for the phase transition[173]. The structural representations that belong to different phases always cluster at different locations in the latent parameter space, as is shown schematically in Fig. 6(b)[174]. Utilizing the decoder, the evolution of data can be directly visualized by continuous tuning of the input latent parameter. For example, the lattice dynamics in monolayer TMDs or the structural transformation in disordered systems has been reconstructed in real space with decoders and agreed well with numerical calculations[174,180]. Moreover, the intermediate activation visualization from the convolutional encoder was found to accurately represent the correlation function of the quantum system[181,182] (e.g., the correlation of the spin states between adjacent lattices). A shift-invariant variational AE has recently been utilized[183] to discover repeated features in STEM and STM images of atomic structure. AE can also be used to automatically detect the new phases of matter and explore the phase diagrams based on anomaly detection, where the large reconstruction loss indicates a new phase of matter that is not included in the original training dataset.[176]. The complex behavior of strongly correlated quantum systems can be visualized in real space and can be spanned into a continuous image vector space with the generative part of AE. Going one step further, a deep AE, fed by experimental SPM images and advanced simulation methods[184,185], can make quantitative comparisons in the latent parameter space and is likely to enable a better understanding of the mechanism for the nanoscale phase composites.

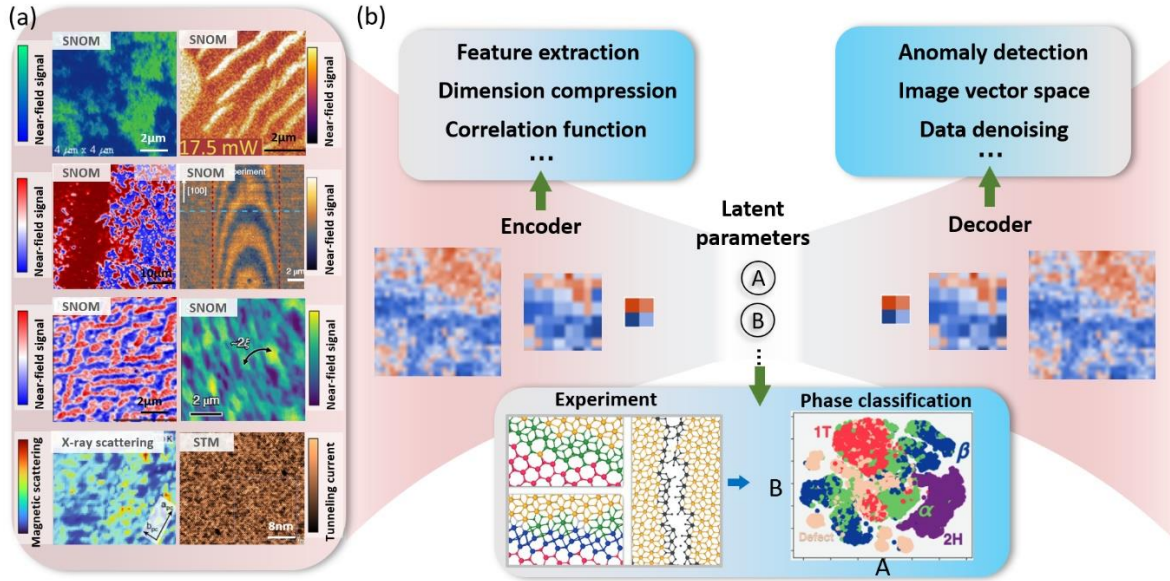

**Figure 5. Proposed unsupervised learning with an autoencoder.** (a) SPM images of quantum matter probed by s-SNOM and other scanning probe methods[55,107,109,161,163,164,169,170]. (b) Unsupervised learning approach for quantum matter image-like data. The network structure for an AE is shown in the center of the diagram. The encoder extracts patterns in the data, compresses the dimension of the images[186], [PRR, 2020. Machine learning holographic mapping by neural network renormalization group] and identifies the correlation function by intermediate activation visualization. The latent parameters, which act as an information bottleneck, fully characterize the data with a few variables and can cluster different phases in the latent parameter space. The decoder of the AE discovers new phases of matter by anomaly detection, generates the new image data in the image vector space in a continuous way, and denoises the data.

**Intelligent scanning local probes**

AI and ML have the potential to eliminate repetitive and time-consuming routines from SPM experiments. Tedious tasks required to identify a specific parameter space for a given experiment can be automated. Here, we outline several algorithms that may enable fast, automatic, and adaptive SPM data acquisition and preprocessing. One promising algorithm often adopted by the ML community is sparse sampling and Gaussian process (GP) regression, which can speed up measurements and reconstruct the signal in unscanned areas with controllable errors. Results are shown in the top panel in Fig.6(a), where the tip spirally scanned 1/6 of the area on the sample surface and the phase separation in ferroelectric domains was well reconstructed with less than 5% error, which provided more accurate reconstruction compared, for example, with compressed sensing[187]. The major prior assumption in the GP is that the signal throughout the image can be characterized by a multivariate Gaussian distribution. The probability distribution (expectation and variance) for the unscanned regions can be derived analytically according to Bayes' theorem for multivariate Gaussians. The data acquisition acceleration enabled by the GP and sparse sampling is ready to be applied to various SPM measurements. In the bottom panel in Fig.5(a), we show the Gaussian regression results for a synthetic polariton wave image, in which the expectation of reconstruction reproduces the image obtained by raster scan. The spatial features of surface polaritons and edge polaritons are perfectly recovered from the sparsely sampled data. It is also demonstrated that the sparse sampling technique can be applied to hyperspectral datasets (Fig. 5(c)[188,189].

A summary of various ML algorithms and their corresponding applications in s-SNOM and other SPM measurements is shown in the right panel in Fig. 6. For example, the optical beam alignment optimization can be automated with the assistance of the machine learning algorithm M-LOOP, and iterative operations for aligning spatial and temporal overlap of optical beams can be exempted[37]. In the data collecting process, one always needs to identify the location of the target area on the sample with nanometer precision, and hence classification and segmentation are required. However, because of large noise, artifacts, and subtle feature contrast, real-time classification and segmentation can be challenging. For example, several previous studies applied the CNN for SPM image classification but could only make predictions based on the whole images[38,190,191]. Moreover, NN-based classification and segmentation typically require a large training dataset, extensive computational power, and fixed input dimension.

Fast and real-time segmentation of SPM images can be realized with the help of ML algorithms that attain a simpler structure than CNN or NN. For example, as is shown schematically in Fig.6(d), the multidomain AFM images can be transferred into feature maps after local Fourier transform and dimension reduction based on Principal Component Analysis (PCA)[192]. The idea of the adaptive scans was also proposed in a recent study, where the phase boundary could be located and classified by analyzing the relative spatial variation of amplitude and phase signal with Support Vector Machine (SVM)[193]. The AFM controller decides the following scanning area based on the SVM output without human assistance. In some cases, the probed signal doesn't directly correspond to the underlying sample properties and further interpretation is required. For example, the CO-AFM (AFM with a CO molecule attached to the metal tip) measurements on 2D molecule systems contained hydrogen and carbon; the complicated tip-sample interaction impeded the interpretation of measured data[194]. Progress with this notoriously complicated inverse imaging problem has been achieved by utilizing image descriptors based on the CNN model, where the atom-type images are extracted from AFM images measured at different signals, as is shown in Fig. 6(e)[158].

We foresee several directions for further improving the SPM measurements via more advanced ML methods. One potential direction is to achieve higher resolution based on existing instruments. By intelligently adding "jumper wire" connections to construct "express" communication pathways among non-adjacent layers, a traditional CNN network can resolve sub-picometer atom defects in STEM images[195]. In addition, a denoising autoencoder can be applied to remove the noise and artifacts in the AFM images[190,191] and the super-resolution optical images can be achieved using a generative adversarial network (GAN), which is presented in Fig. 6(f)[196].

Real-time control of microscopy measurements is of great interest to the goal of optimizing data acquisition time, which currently can be several days. In atom-resolved microscopies such as STM, human operation can be replaced with a deep-learning workflow for feature detection and thus with automated experiments using the ensemble of learning-iterative training[50,54]. Further, the tip conditioning in scanning tunneling microscopy can be analyzed using trained CNN methods for higher efficiency[30]. Lastly, the experimental observations can be compared with the theoretical models via Bayesian optimization in a quantitative way[197].

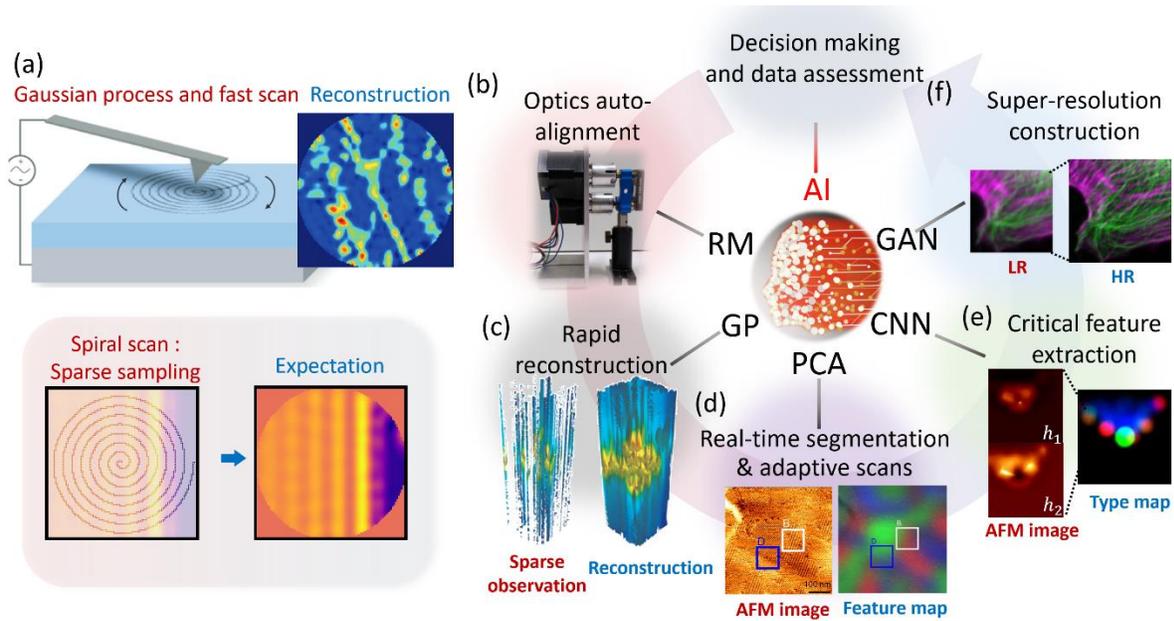

**Figure 6. Smart SPM and REAL-SNOM (REinforced & Advanced Learning SNOM).** (a) Top panel: Schematics of spiral scan and Gaussian process[187]. Bottom panel: Spiral scan and Gaussian process for polariton images. (b) Automated laser beam alignment based on Raspberry Pi auto-aligner[37]. c) Rapid reconstruction based on the Gaussian process for hyperspectral SPM data[188]. (d) Different domains in the AFM images are mapped in the feature map in real time based on principal component analysis[192]. (e) The types of atoms are extracted and presented on the type map from the CO-AFM images[158]. (f) Super-resolution construction of optical images based on generative adversarial networks[196]. AI: artificial intelligence; GAN: generative adversarial network; CNN: convolutional neural network; PCA: principal component analysis; GP: Gaussian process; RL: reinforcement learning. (b) – (f) are specifically relevant to REAL-SNOM.

**Outlook and Discussion: REinforced & Advanced Learning (REAL) SNOM**

The ML algorithms mentioned above have the potential to significantly enhance the performance of s-SNOM and SPM techniques in general. It is important to realize that the emergence of ML for s-SNOM analysis is only recent[32,33]. This is still largely unexplored territory, where interesting advances are bound to occur in the near future, propelled in part by the large volumes of data generated in user facilities as well as in university labs across the globe. We propose the blueprint for a future REinforced & Advanced Learning SNOM (REAL-SNOM) that will partially or fully automate data collection, conditioning, data assessment, and decision making. Fast and adaptive scanning can be realized by taking advantage of sparse sampling, reconstruction algorithms, and real-time segmentation. The state of the tip can be evaluated in situ by analyzing the SPM images with neural networks[30,54,198]. The deep kernel learning (DKL)[199], which can be viewed as a combination of NN and GP kernels, predicts the signal and its uncertainty in the unscanned region[200]. The REAL-SNOM could actively select future scanning areas based on the DKL output and the pre-defined acquisition function[201,202]. After establishing the system for decision-making, the REAL-SNOM selects actions based on the environmental information to maximize the reward scores, which will ultimately outperform experienced human experimentalists.

It is prudent to point out that ML-based analysis is prone to certain sources of bias that may be difficult to eliminate in practice[1,203–205]. In a complex system such as an optical scanning probe experiment, the number of parameters relevant to the outcomes of measurement is often intractable. For example, variations in tip geometry such as tip length, apex radius, shape, and even minor wear and tear due to fabrication accuracy and experimental conditions are all relevant factors that should be considered. Not accounting for those factors explicitly could lead to systematic bias. Furthermore, the implementation of certain algorithms or ML models in connection to realistic experiments may require knowledge of the experimental conditions and the samples. For example, the GP reconstruction from spiral scan data requires scale matching between the spatial features of the samples and the density of the spirals. Initial guesses and attempts might be needed when the algorithms are performed on new and unknown systems. Other deeper issues such as algorithmic biases are all important aspects of the modeling that require great attention and further studies[206,207]. With AI/ML-assisted data processing, reconstruction and augmentation quickly becoming ubiquitous, research communities will need to establish norms for responsible reporting of the measurements. An incomplete list of already pressing issues includes the following: Researchers need to clearly identify actually measured data and reconstructed (parts of) data/images. If the trained network analysis is involved, sufficient details of training sets need to be reported. Furthermore, it is imperative to establish the minimum test requirements to validate the robustness of network-derived results. In short, how do we keep in check the "black box" aspects of AI/ML-assisted outcomes of measurements and analysis?